\input{aipcheck.tex}
\documentclass[final]{aipproc}
\layoutstyle{6x9}

\def\be{\begin{equation}}
\def\ee{\end{equation}}
\def\bea{\begin{eqnarray}}
\def\eea{\end{eqnarray}}
\def\ba{\begin{array}}
\def\ea{\end{array}}
\newcommand{\xbf}[1]{\mbox{\boldmath $#1$}}

\begin{document}

\title[]{Charge radii of octet and decuplet baryons}
\author{Neetika Sharma}{address={Department of Physics, Dr. B.R.
Ambedkar National Institute of Technology, Jalandhar, Punjab-144
011, India}}
\author{Harleen Dahiya}{address={Department of Physics, Dr. B.R.
Ambedkar National Institute of Technology, Jalandhar, Punjab-144
011, India}}
\date{\today}

\begin{abstract}
The charge radii of the octet and decuplet baryons have been
calculated in the framework of chiral constituent quark model
($\chi$CQM) using a general parameterization (GP) method. Our
results are comparable with the latest experimental studies as well
as with other phenomenological models. The effects of SU(3) symmetry breaking and GP
parameters pertaining to the one-, two- and three-quark
contributions have also been investigated in detail.
\end{abstract}
\maketitle

The internal structure of baryons is determined in terms of
electromagnetic Dirac and Pauli from factors $F_1(Q^2)$ and
$F_2(Q^2)$ or equivalently in terms of the electric and magnetic
Sachs form factors $G_E(Q^2)$ and $G_M(Q^2)$ \cite{sach}. The
electromagnetic form factors
are further related to the static low energy observables of charge
radii and magnetic moments. Although Quantum Chromodynamics (QCD) is
accepted as the fundamental theory of strong interactions, the
direct prediction of these kind of observables from the first
principle still remains a theoretical challenge as they lie in the
nonperturbative regime of QCD.

The mean square charge radius ($r^2_B$), giving the possible
``size'' of baryon, has been investigated theoretically in various models such as Skyrme model \cite{kunj}, $1/N_c$ expansion \cite{buch07}, chiral perturbation
theory \cite{pcqm}, lattice QCD \cite{wang} etc. The results for
different theoretical models are however not consistent with each
other. Several measurements have been also made for the charge
radii of $p$, $n$, and strange baryon ${\Sigma^-}$
\cite{pdg,sigma-}.

The chiral constituent quark model ($\chi$CQM) \cite{manohar} coupled with the ``quark sea'' generation through the chiral fluctuation of a constituent quark into Goldstone bosons (GBs) \cite{cheng,hd}, finds applications in the low energy regime.
Since this model successfully explains many of the low energy hadronic matrix elements \cite{hdmagnetic,hds,nsweak,hdcharm}, it therefore become desirable to extend it to calculate the charge radii of the octet and decuplet baryons using a general parametrization method (GP) \cite{morp89}.

The most general form of charge radii operator consisting of the sum of one-, two-, and three-quark terms with coefficients ${\mathrm A}$, ${\mathrm B}$, and ${\mathrm C}$ is expressed as \be
\widehat{r^2} = {\mathrm A} \sum_{i=1}^3 e_i {\bf 1} + {\mathrm B}
\sum_{i \ne j}^3 e_i \, {\xbf{\sigma}_i} \cdot \xbf{\sigma}_j +
{\mathrm C} \sum_{i \ne j \ne k }^3 e_i \, \xbf{\sigma}_j \cdot
\xbf{\sigma}_k \,. \label{rad} \ee Solving the charge radii
operators for the spin $\frac{1}{2}^+$ and spin $\frac{3}{2}^+$
baryons, we obtain \bea \widehat{r^2_{B}} &=& ({{\mathrm A}}-
3{\mathrm B})\sum_i e_i + 3( {\mathrm B} - {\mathrm C}) \sum_i e_i
\sigma_{i z} \,, \label{r1/2}
\\ \widehat{ r^2_{B*}} &=& ({\mathrm A} - 3{\mathrm B} + 6 {\mathrm
C}) \sum_i e_i + 5( {\mathrm B} - {\mathrm C}) \sum_{i} e_i
\xbf{\sigma_{iz}} \,.\label{r3/2} \eea The charge radii squared
${r^2_{B(B^*)}}$ for the octet (decuplet) baryons can now be
calculated by evaluating matrix elements corresponding to the
operators in Eqs. (\ref{r1/2}) and (\ref{r3/2}) and are given as $
r^2_B = \langle B |\widehat{ r^2_B}| B \rangle \,, r^2_{B^*} =
\langle B^* |\widehat{r^2_{B^*}}| B^* \rangle\,,$ where $|B \rangle$
and $|B^* \rangle$ respectively denote the SU(6) spin-flavor
wavefunctions for the spin $\frac{1}{2}^+$ octet and the spin
$\frac{3}{2}^+$ decuplet baryons. Using the ${\rm SU}(6)$
spin-flavor wavefunctions of the octet and decuplet baryons in Eqs.
(\ref{r1/2}) and (\ref{r3/2}) as well as the $\chi$CQM parameters, $a$, $a
\alpha^2$, $a \beta^2$, and $a \zeta^2$ representing the
probabilities of fluctuations to pions, $K$, $\eta$, and $\eta^{'}$,
respectively, the charge radii of other octet and decuplet baryons
can be calculated. The results have been presented in Table \ref{chargeradii1}.

To understand the implications of chiral symmetry breaking and
``quark sea'', we have also presented the results of NQM including
the one-, two-, and three-quark contributions of the GP parameters.
If we consider the contribution coming from one-quark term only, the
charge radii of the charged baryons are equal whereas all neutral
baryons have zero charge radii in NQM. These predictions are
modified on the inclusion of two- and three- quark terms of GP
method in NQM and are further modified on the inclusion of ``quark
sea'' and SU(3) symmetry breaking effects. Thus, it seems that the
GP parameters alone are able to explain the experimentally observed
non-zero charge radii of the neutral baryons. However, NQM being unable
to account for the ``proton spin problem'' and other related
quantities, the results have been presented for $\chi$CQM.

The importance of strange quark mass has been investigated by comparing
the $\chi$CQM results with and without SU(3) symmetry breaking. The
SU(3) symmetry results can be easily derived by considering
$\alpha=\beta=1$ and $\zeta = -1$. The SU(3) breaking results are in
general higher in magnitude than the SU(3) symmetric results. The
SU(3) symmetry breaking corrections are of the order of 5\% for the
case of $p$, $\Sigma^+$, $\Sigma^-$, and $\Xi^-$ baryons whereas
this contribution is more than 20\% for the neutral octet baryons.
We have also compared our results with the other phenomenological
models and our results are in fair agreement in sign and magnitude
with the other model predictions. Since experimental information is
not available for some of these charge radii, the accuracy of these
relations can be tested by the future experiments.

The decuplet baryon charge radii, presented in Table
\ref{chargeradii1}, the inclusion of SU(3) symmetry breaking
increases the predictions of charge radii as in case of octet
baryons. Again, the sign and magnitude of the decuplet baryon charge
radii in $\chi$CQM are in fair agreement with the other
phenomenological models with the exception for neutral baryons. One
of the important predictions in $\chi$CQM is a non-zero $\Delta^0$
charge radii which vanishes in NQM as well as in some other models.
The contribution of the three-quark term in the case of decuplet
baryons is exactly opposite to that for the octet baryons. Unlike
the octet baryon case, the inclusion of the three-quark term
increases the value of the baryon charge radii.

The $\chi$CQM  using a GP method is able to provide a fairly good
description of the charge radii of the octet and decuplet baryons.
The most significant prediction of the model is the non-zero value
pertaining to the charge radii of the neutral baryons. The SU(3)
symmetry breaking parameters pertaining to the strangeness
contribution and the GP parameters pertaining to the one-, two- and
three-quark contributions are the key in understanding the octet and
decuplet baryon charge radii. New experiments aimed at measuring the
charge radii of the other baryons are needed for a profound
understanding of the hadron structure in the nonperturbative regime
of QCD. Thus at the leading order constituent quarks and the weakly
interacting Goldstone bosons constitute the appropriate degrees of
freedom in the nonperturbative regime of QCD.
\section{ACKNOWLEDGMENTS}
N.S. would like to thank CSIR and the organizers of Baryons'10 for
financial support. This work is a part of DAE-BRNS project (Grant
No. 2010/37P/48/BRNS/1445).

\begin{table}
{
\begin{tabular}{|c|c|c |c|c|c|c| c|c|c| }\hline

Charge & NQM & \multicolumn{3}{c|}{$\chi$CQM$_{{\rm config}}$ }&
Charge & NQM & \multicolumn{3}{c|}{$\chi$CQM} \\\cline{3-5}
\cline{8-10}
radii & & with SU(3) & \multicolumn{2|}{c}{with SU(3)}
& radii & & with SU(3) & \multicolumn{2|}{c}{with SU(3)} \\
&&  symmetry & \multicolumn{2}{c|}{symmetry breaking}&& & symmetry &
\multicolumn{2}{c|}{symmetry breaking} \\ \cline{3-5} \cline{8-10}
&&& $ {\mathrm A} = 0.879 $ & $ {\mathrm A} = 0.879$
& && & $ {\mathrm A} = 0.879 $ & $ {\mathrm A} = 0.879$ \\
& && $ {\mathrm B} = 0.094 $ & $ {\mathrm B} = 0.094$ && && $
{\mathrm B} = 0.094 $ & $ {\mathrm B} =
0.094$\\
& && $ {\mathrm C} = 0.0 $ & $ {\mathrm C} = 0.016$&& &&
$ {\mathrm C} = 0.0 $ & $ {\mathrm C} = 0.016$ \\ \hline

& && && ${r^2}_{\Delta^{++}}$ &1.084 &0.938& 0.961 & 0.996\\
$r^2_p$  &0.813   &0.732 & 0.801 & 0.766
&${r^2}_{\Delta^{+}}$ &1.084&0.938& 0.946 & 0.983 \\
($r_p$= 0.877$\pm$ 0.007 \cite{pdg}) & && &&  & & &  & \\

$r^2_n$  &$-$0.138 &$-$0.087 &$-$0.140
&$-$0.116 & ${r^2}_{\Delta^{0}}$ &0.0&0.0&
$-$0.030 & $-$0.025 \\
($-$0.1161 $\pm$ 0.0022 \cite{pdg}) & && &&  & & &  & \\

& && && ${r^2}_{\Delta^{-}}$ &1.084&0.938& 1.006 & 1.033 \\
$r^2_{\Sigma^+}$ & 0.813 &0.732 &0.802 &0.767 &
${r^2}_{\Sigma^{*+}}$ &1.084& 0.938& 0.940 & 0.978 \\

$r^2_{\Sigma^-}$  &0.675 &0.646 &
0.678 &0.664& ${r^2}_{\Sigma^{*-}}$ &1.084&0.938 & 1.013 & 1.038 \\
(0.61 $\pm$0.21 \cite{sigma-}) & && &&  & & &  & \\

$r^2_{\Sigma^0}$ &0.069 &0.043 & 0.062 &0.052 &
${r^2}_{\Sigma^{*0}}$ &0.0&0.0 &$-$0.036 & $-$0.030 \\
$r^2_{\Xi^0}$ &$-$0.138 &$-$0.087 &$-$0.145&$-$0.120&
${r^2}_{\Xi^{*0}}$ &0.0 &0.0 &$-$0.043 & $-$0.035 \\
$r^2_{\Xi^-}$ & 0.675 &0.646 & 0.683 & 0.669& ${r^2}_{\Xi^{*-}}$
&1.084&0.938 & 1.019 & 1.043 \\
$r^2_{\Lambda}$ &$-$0.069& $-$0.042 & $-$0.076 &$-$0.063 &
${r^2}_{\Omega^-}$ &0.390&0.245 & 0.429 & 0.355 \\
\hline
\end{tabular}
\caption{Charge radii of octet and decuplet baryons in $\chi$CQM (in
units of fm$^2$).} \label{chargeradii1}}
\end{table}

\end{document}